\documentstyle[multicol,prl,aps,epsf,psfig,epsfig]{revtex}
\begin{document}
\title
{
Time evolution of link length distribution in PRL collaboration network
}
\author
{
  Parongama Sen, Anjan Kumar Chandra, Kamalika Basu Hajra and Pratap Kumar Das 
}
\address
{
 Department of Physics, University of Calcutta,
    92 Acharya Prafulla Chandra Road, Kolkata 700009, India.
}
\maketitle
\begin{abstract}

An important aspect of a Euclidean network is its link length distribution,
studied in a  few  real networks so far. We compute the  distribution
of the link lengths between collaborators whose papers appear in the Physical
Review Letters (PRL) in several years within a range of four decades. 
The distribution is non-monotonic; there is a peak at 
nearest neighbour distances followed by a sharp fall and a subsequent rise
at larger distances. The behaviour of the statistical properties of the 
distribution with time indicates that collaborations might become distance
independent in about thirty to forty years.

%

\end{abstract}

Preprint number CUPhysics20/2005
\begin{multicols}{2}

Ever since the discovery of small world effect in a variety
of networks \cite{watts}, study of real world networks 
and their theoretical modelling have generated tremendous activity.
A network is equivalent to a graph and is characterised by the links which
connect pairs of nodes. 
Based on observations and theoretical arguments, it has been 
established that factors like preferential attachment, duplication,
aging etc. are responsible in determining the connectivity in 
many real world networks \cite{BA}.

In a Euclidean network, where the nodes are embedded on a Euclidean space, it 
can be expected that the distance between nodes will play an important role 
in determining whether a link will  connect them. In several
theoretical models of Euclidean network, the link length distribution 
has been assumed to have a power law decay \cite{ps_rev}.  


Linking schemes in a  few real world networks  in which 
geographical distance plays an important have been studied. 
These are the internet \cite{yook}, transport \cite{newman_ll}, neural network \cite{eguiluz}
and some collaboration networks \cite{katz,nagpaul,drift,olson}.
In this article, we report the study of a network of collaborators
whose papers appear in Physical Review Letters.
We also study this distribution at different times as it is a 
dynamic network and reflects  the evolution of both  communication and
human interactions.

Scientific collaboration network is a social network \cite{newman1,newman2}
in which close
personal encounters are essential to a large extent and it is 
expected that the existence of links between authors will depend on the
distance separating them. Communication is the key factor in 
a collaboration and it has undergone revolutionary changes over the 
years.  This effect will
 manifest in the time evolution of the link length distribution. 
We have therefore studied its behaviour  over four 
decades.

To obtain the link length distribution, one should take the 
collaboration network and
calculate the geographical distances separating the host institutes of the authors
who share a link. However, this becomes a formidable task.
We have obtained the distance distribution in an indirect way. 
Noting that the collaboration acts are the papers, the distance between the
co-authors in a particular paper would also supply the necessary data.
We have therefore taken sample papers (at least 200 for each year) from the 
Physical Review Letters (PRL)  and  
 calculated the geographical distance between each pair of 
authors in a coarse grained manner for nine  different years between 1965 to 2005
and obtained the link-length distributions. 

The pair-wise distances $l$ gives the distribution $P(l)$ of
the distance between two collaborating authors.
We have also defined a distance factor $d$ for each paper where 
$d$ is the average of the pair-wise distances of authors coauthoring 
that paper. The corresponding  distribution $Q(d)$ has also been computed.
For example, let there be a paper authored by three scientists and
let $l_{12}, l_{13}, l_{23}$ be the pairwise distances.
Then
$d = (l_{12}+ l_{13}+ l_{23})/3$. Note that in $P(l)$, the fact that 
$l_{12}, l_{13}$ and $l_{23}$ are obtained from a single collaboration act 
is missing. Hence, in a sense, $Q(d)$ takes care of the correlation between the
distances. Let us call $Q(d)$ the correlated distance distribution.

In principle, the actual geographical distances have to be 
computed which is  non-trivial. We have coarse 
grained the
distances in a convenient way. To  author X in a paper we associate
the indices $x_1, x_2, x_3$ and $x_4$ ($x_i$'s are integers) which 
represent the 
University/Institute, city, country and continent of X respectively. 
Similar
indices $y_1, y_2, y_3$ and $y_4$ are defined for author Y. If, for
example, authors X and Y belong to the same institute, $x_i=y_i=1$ for
all $i$. On the other hand, if they are from different countries but
the from same continent, $x_4=y_4$ but $x_i \neq y_i$ for $i <  4$.  
We find out for what
maximum value of $k$,  $x_k \ne y_k$. The distance  between X and Y 
is then $l_{XY} = k +1$. If $x_i = y_i$ for all values of $i$ it means 
$l_{XY} =1$ according to our definition. 
As an example, one may consider the paper  PRL {\bf 64}  2870 (1990),
which features 4 authors. Here authors 1 and 2 are from the same 
institute in Calcutta, India, and are assigned  the variables 1, 1, 1, 1.
The 3rd  author belongs to a different institute in Calcutta and therefore
gets the indices 2, 1, 1, 1. The last author is from an institute in Bombay, India, 
and is assigned the variables 3, 2, 1, 1. 
Hence $l_{12} = 1, l_{13}=l_{23}=2, l_{14}=l_{24}=l_{34}=3$ and the
average $d=2.333$.
Defining the distances in this way, the values of $l$ are 
discrete while the $d$ values have a continuous variation.
For papers with two authors, the two distributions are identical
but will be different in general.

\begin{center}
\begin{figure}
\noindent \includegraphics[clip,width= 5cm, angle=270]{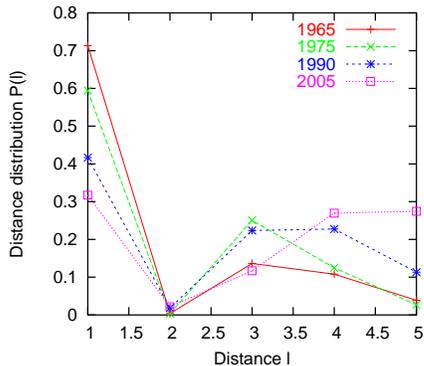}
\caption{Distance distribution $P(l)$ as function of distance $l$ for
              different years. } 
\end{figure}
\end{center}
\begin{center}
\begin{figure}
\noindent \includegraphics[clip,width= 5cm, angle=270]{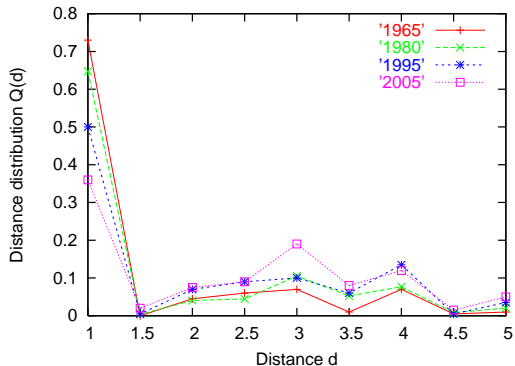}
\caption{Correlated distance distribution $Q(d)$ vs distance $d$ plot for 
different years are shown.}
\end{figure}
\end{center}

We have made exception for USA authors since it is a big country
comparable in size to Europe which consists of many countries. 
 Thus  two authors belonging to, say, Kentucky and Maryland will have
 different country indices, i.e., 
 $x_3 \ne y_3$.

 Some papers like the experimental high energy 
 physics ones typically involve many authors and many institutes. We 
 have considered an upper bound, equal to 20, to
 the number of institutes and no bounds for the number of authors.
 In case of multiple addresses, only the first one has been
 considered.

Both the distributions  $P(l)$ and $Q(d)$ have the following features:\\
1. A peak  at  $l$ or $d =1$\\
2. A sharp fall at around $l$ or $d=2$ and a subsequent rise.
The fall becomes less steep in time.\\
3. Even for the most recent data, the peak at nearest neighbour distances 
is quite dominant. However, with the passage of time, the peak value  
at nearest neighbour distances shrinks while 
the probability at larger distances increases.

In Figs. 1 and 2, the distributions $P(l)$ and $Q(d)$ are shown. 
The two distributions have similar features  
but  differ in magnitude, more so in recent years, when the  
number of authors 
is significantly different from two in many papers.
The data for $Q(d)$  apparently  has an oscillatory nature for
larger values of $d$.  
However, we believe that these oscillations are due to the coarse graining 
of the data and it is more likely that the peak at the nearest neighbour 
distances is followed by a crest and a gentle hump at larger distances.
The hump  grows in size with time while the peak value at
nearest neighbour distances diminishes.
\begin{center}
\begin{figure}
\noindent \includegraphics[clip,width= 5cm, angle=270]{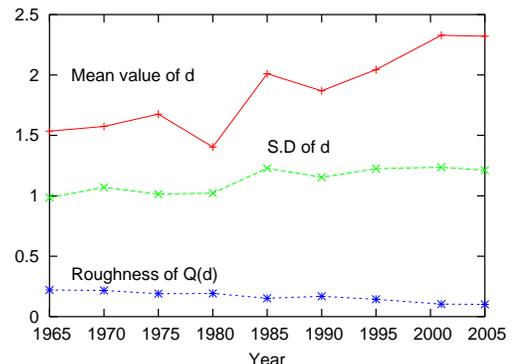}
\caption{ The mean value and standard deviation of
                distances $d$ increase with time while the 
                roughness of the distance distribution $Q(d)$  shows a steady decrease.}
\end{figure}
\end{center}

We make a detailed analysis of $Q(d)$, the correlated distance distribution.
In Fig. 3, we present the results.
The mean increases appreciably  in consistency with 
our idea that with the progress of time there will be more collaborations
involving people working at a distance.
The fluctuation  also shows an increase, although its increase
is not that remarkable since the total range of interaction remains fixed in our convention. 
If collaborations were really distance independent, the distributions
$Q(d)$ and $P(l)$ would have looked flat. We have estimated the deviation of $Q(d)$
from a flat distribution by calculating its ``roughness'' $R_Q$  defined as 
$\sqrt{ \langle(Q(d) -\bar Q(d))^2\rangle} $ where $\bar Q(d)$ is the mean value of 
$Q(d)$.
$R_Q$ shows a  decrease with time which is approximately linear. 

The above results imply that even with the communication
revolution, most
collaborations take place among nearest geographical neighbours.
The drop near $d=2$ maybe justified from the fact that in most
cities one has only one university/institute and when one 
collaborates with an outsider, she or he belongs to some other 
city or  country in most cases.
There is some indication that in the not too distant future
collaborations will become almost distance independent as in Fig. 3, 
$R_Q$ seems 
to vanish at around 2040 when extrapolated.
It may also happen that $R_Q$ saturates to a finite value in the coming years,
and perhaps it is  too early to predict anything definite.

What is the nature of the distribution when the real distances are considered?
We notice that there is a sharp decrease of $Q(d)$ with $d$ 
initially which may be assumed to be exponential in nature. 
The way we have defined $l$ (or $d$), it  maybe assumed that the true  
distances $d_{real}$ scale roughly as $\exp (\alpha d^a)$ where $a$ is a number of
the order of unity.
In that case, the initial exponential decay of $Q(d)$ with $d$ 
corresponds to a power law decrease with $d_{real}$. 
The subsequent rise of the distribution with $d$ should also show up
against  $d_{real}$.

In summary, we have studied the link length distributions in the Euclidean
network of collaborators of PRL papers. Unlike the other features of a network, 
e.g., degree
distribution or aging,  
 we do not find a  conventional power law or exponential decay but rather a 
non-monotonic behaviour. 
The data over different times shows that the communication revolution has indeed
influenced long distance collaborations to a considerable extent

Acknowledgments: We thank
the fellow members of the network group of Calcutta 
for helpful discussions.
We   acknowledge support from CSIR grants  03(1029)/05/EMR-II (PS),
9/28(609)/2003-EMR-I (PKD) and
9/28(608)/2003-EMR-I (KBH).
AKC is grateful to support from UGC.

\end{multicols}
\end{document}